\documentclass[cameraready]{Interspeech}

\usepackage{times}
\usepackage{latexsym}
\usepackage{hyperref}
\usepackage{url}
\usepackage{appendix}
\usepackage{amsmath,graphicx,booktabs,multirow,tabularx}
\usepackage{wrapfig}
\usepackage{makecell}
\usepackage{microtype}

\title{SCNet: Enhancing GAN-based Speech Generation with Subband Condition Network and Magnitude-aware Phase Loss}

\author[affiliation={1}]{Nan}{Xu}
\author[affiliation={2}]{Mingxue}{Yang}

\address{
    $^1$ WeChat, Tencent Inc, Beijing, China  \\
    $^2$ University of Electronic Science and Technology of China
}

\email{vicnxu@tencent.com, michelleyang2017a@gmail.com}


\usepackage{comment}

\begin{document}

\maketitle

\begin{abstract}
Recent speech generation has been predominantly driven by GAN-based networks aimed at high-quality waveform synthesis from mel-spectrograms. However, these methods often operate as black-box models, leading to the loss of inherent spectral information. In this work, we propose SCNet, a GAN-based vocoder augmented with a Subband Condition Network to address this issue. Specifically, SCNet leverages a subband signal predicted by a lightweight condition network as prior knowledge. This subband signal is then transformed via STFT to obtain Fourier coefficients, which are integrated into the backbone for the enhanced reconstruction. Additionally, to mitigate the phase wrapping, we introduce a magnitude-aware phase loss that computes instantaneous phase errors weighted by the corresponding magnitude, emphasizing regions with higher energy. Experimental results demonstrate that SCNet achieves superior performance in both objective and subjective evaluations for high-quality speech generation.

\end{abstract}

\section{Introduction}
\sloppy
In the real world, speech is an extremely important modality for various practical applications. Neural network based vocoders aiming to generate the high-quality waveform from an intermediate representation play a crucial role in speech or audio synthesis \cite{hu2025chain,xu2025universal,he2025ritta}. In particular, mel-spectrogram representations which have approximate human auditory perceptions and compact dimensionality are widely used as the intermediate representations, especially in text-to-speech (TTS) \cite{ren2019fastspeech,du2024cosyvoice,meng2025autoregressive,chen2025f5}, singing voice synthesis (SVS) \cite{liu2022diffsinger,lei2023unisyn,zhang2024stylesinger} and voice conversion (VC) \cite{qian2019autovc,yao2024promptvc,yuguang2025takin,zuo2025rhythm} technologies. A two-stage strategy is always used in these methods: the intermediate mel-spectrogram representation is first predicted from source feature and next stage converts it into a raw waveform. The traditional signal processing approaches mainly focus to map intermediate feature to the original speech, which introduces nonnegligible artifacts. In recent years, with the success of deep learning, mel-spectrogram based neural vocoders have been rapidly improved in the aspect of quality and naturalness of speech.

\sloppy
Generative adversarial network based neural vocoders are one family of methods that are the most effective due to the high-quality waveform generation and fast inference speed. GAN-based vocoders are usually driven by two major categories: direct waveform generation \cite{kumar2019melgan,kong2020hifi,jang2021univnet,bak2023avocodo,leebigvgan,shen2024fa} and inverse Short-Time Fourier Transform (iSTFT) based methods \cite{ai2020neural,kaneko2022istftnet,kaneko2023istftnet2,ai2023apnet,du2023apnet2,siuzdak2023vocos}. The former usually operates in the time-domain, where temporal transposed convolution modules are utilized to transform the mel-spectrogram to the raw waveform by directly sequential upsample processes. In contrast, iSTFT-based methods predict magnitude and phase spectrums and employ the iSTFT to generate the waveform. These methods typically generate waveform in a black box and have no guidance of the initial condition, leading to fine-grained information loss of the predicted spectrum. During the training process, the feature matching loss is thereby unstable or even gradually increases\footnote{\url{https://github.com/jik876/hifi-gan/issues/59}}. We also present this issue in the experiment section. Some researchers utilize the neural source-filter (NSF) \cite{wang2019neural} to predict the waveform as the prior knowledge to achieve fine-grained waveform generation\footnote{\url{https://github.com/nii-yamagishilab/project-NN-Pytorch-scripts}} \cite{li2023snakegan,li2023hiftnet}. However, existing pitch tracking methods (such as DIO \cite{morise2009fast} and pYIN \cite{mauch2014pyin}) or pre-trained fundamental frequency (F0) estimation networks \cite{kum2019joint} usually yield errors like incorrect voiced/unvoiced flags and pitch halving/doubling \cite{hirst2021measuring,he2024prosodyfm}, resulting in the suboptimal‌ performance of vocoders.

\begin{figure}[t]
  \centering
  \includegraphics[width=0.49\linewidth]{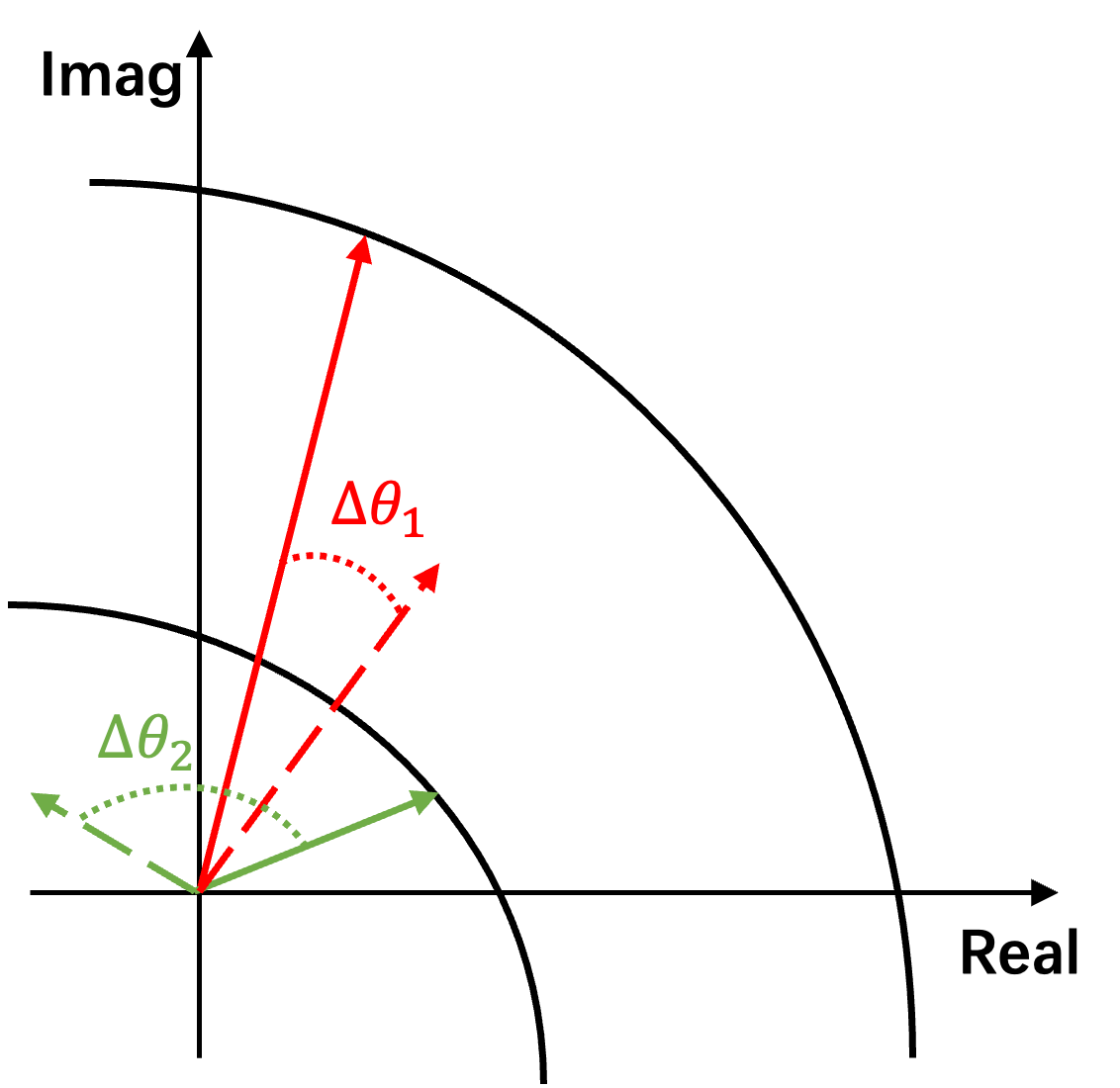}
\caption{Illustration of two possible phase errors in the spectrogram. Red and green colors denote two time-frequency bins. Solid and dash arrows are target and predicted values, respectively. Obviously, the phase error $ \Delta\theta_2 $ is larger than $ \Delta\theta_1 $, leading to a larger weight during training process.}
\label{ss}
\end{figure}

\sloppy
Furthermore, GAN-based vocoders frequently model the magnitude distribution and neglect the inherent phase information. The main reason is that phase distribution has the intricate nature and suffers from the wrapping issue. While iSTFT-based vocoders \cite{kaneko2022istftnet,siuzdak2023vocos} predict the unwrapping phase information by generator networks, the process is still operated in a black box without explicitly supervision and further impairs the accuracy of phase prediction. Additionally, researchers have also explored designs of diverse phase losses for phase modeling \cite{ai2023apnet,du2023apnet2,ai2023neural}. These phase losses generally have intricate expressions and equally consider phase errors of all time-frequency bins. Notably, the magnitude spectrogram is considered as a large proportion of speech perceived quality \cite{gerkmann2015phase}. Nevertheless, the small phase error with large magnitude in Figure~\ref{ss} is considered as less attention, which degrades the overall performance.

\sloppy
To address the aforementioned challenges, we propose the {\bf SCNet}, a novel GAN-based vocoder with the {\bf S}ubband {\bf C}ondition {\bf Net}work and magnitude-aware anti-wrapping phase loss. Specifically in our work, we adopt the iSTFTNet \cite{kaneko2022istftnet} as our backbone network due to its speed efficiency and high synthesized speech quality. To generate the fine-grained spectrogram, we design a subband condition network called CondNet to leverage inherent information (e.g. real and imaginary) existing in a mel-spectrogram. This prior knowledge is integrated into the backbone network. Notably, instead of generating the full-band signal, we predict a subband signal of low-frequency domain. The generation of the full-band signal usually requires the larger frame shift, which suffers from the deterioration of phase continuity \cite{du2023apnet2,ai2023long}. Additionally, to avoid the phase wrapping problem, we design a periodic phase loss, where the square of sine function is used to calculate the difference between predicted and raw phases. Raw magnitude values of each time-frequency unit are assigned as weights for phase loss calculation. Specifically,  the main contributions in our paper are as follows:

\sloppy
\begin{itemize}
\item We propose SCNet, a dual-branch GAN-based vocoder, trained to generate the raw waveform with the frequency-domain prior information. A subband condition network, termed CondNet, is used to predict Fourier spectral coefficients. Unlike previous iSTFT-based models that rely on full-band spectral coefficients, CondNet only predicts subband information of the low-frequency domain, contributing to better synthesized speech quality.
\item To further improve the speech fidelity, a novel magnitude-aware anti-wrapping phase loss is proposed. The phase issue is effectively avoided due to the even function and periodicity of the proposed phase loss. Additionally, we use the raw magnitude values to achieve adaptive assignments of phase loss weights.
\item Our extensive experiment results demonstrate that SCNet matches superior speech quality in terms of subjective and objective metrics. Additionally, we also validate the effectiveness of the proposed CondNet and anti-wrapping phase loss. Furthermore, our SCNet achieves the competitive inference speed compared to other baseline methods especially conventional GAN-based vocoders.
\end{itemize}

\sloppy
The rest of our proposed paper is organized as follows: In Section~\ref{work}, some related vocoder research methods are introduced. Next in Section~\ref{format}, we introduce the proposed dual-branch method, including the subband condition network architecture and the magnitude-aware anti-wrapping phase loss. Experimental results are reported in Section~\ref{exp}. Finally, we present the limitation and conclusion of our paper

\begin{figure*}[t]
\begin{minipage}[b]{1.0\linewidth}
  \centering
  \centerline{\includegraphics[width=13.5cm]{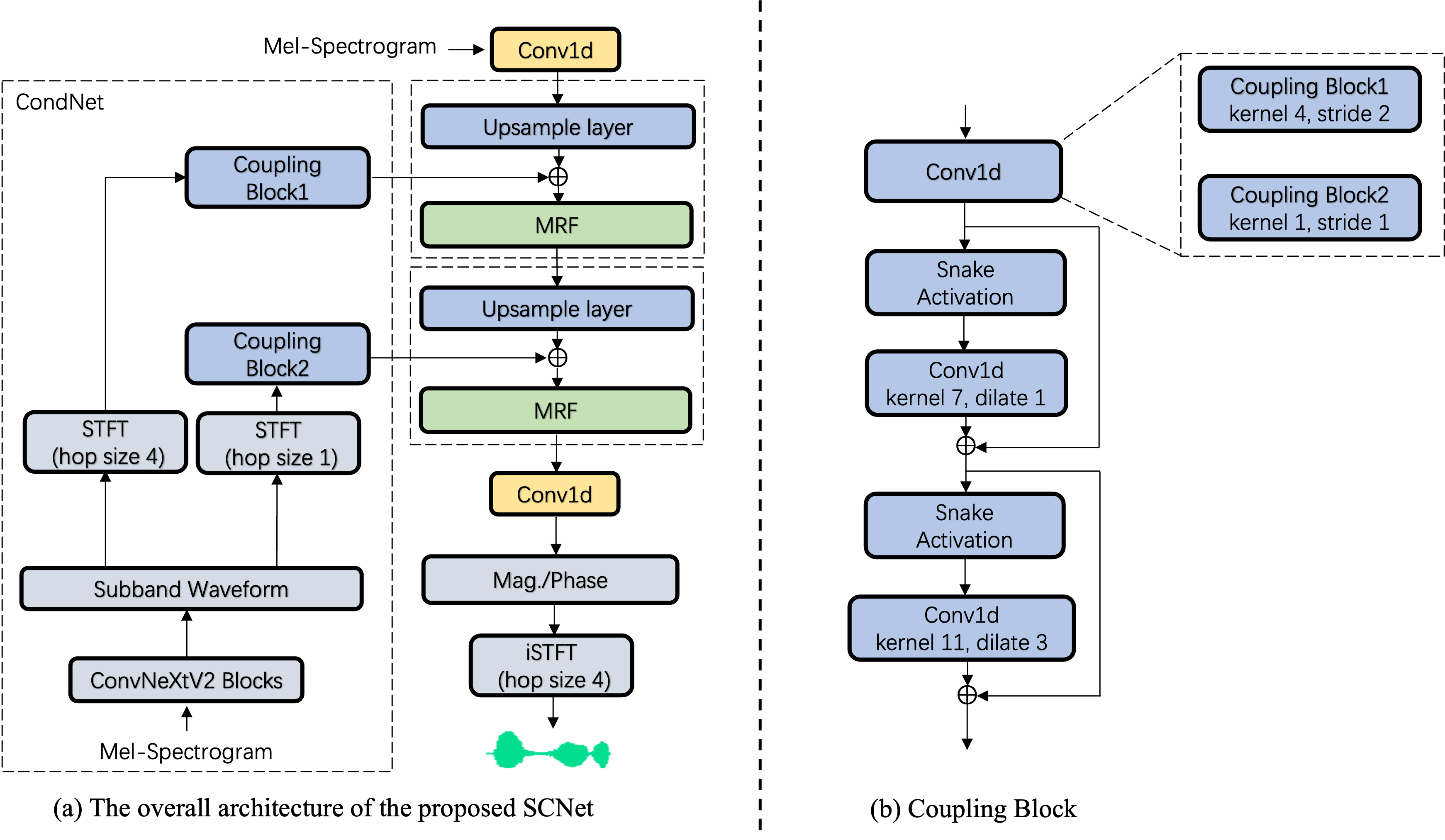}}
\end{minipage}
\caption{The overall architecture of the proposed SCNet. In the subfigure (a), the CondNet architecture is shown in the left and ConvNeXtV2 blocks are used to predict the magnitude and phase in a low frequency subband and further generate the subband waveform. Then, two coupling blocks that take the corresponding STFT output as input are utilized for condition integrations. Notably, MRF denotes the Multi-Receptive Field Fusion module. In the subfigure (b), the coupling block is shown. Other than the first convolution layer, the network configurations are the same for different coupling blocks.}
\label{frmw}
\end{figure*}

\section{Related Work}
\label{work}
\noindent{\bf Direct waveform generation.} Compared to conventional vocoders, GAN-based direct generation methods are gaining growing attention due to their efficient ability for waveform generation. MelGAN \cite{kumar2019melgan} utilizes a non-autoregressive fully convolutional feed-forward architecture for waveform generation without additional perceptual loss functions. HiFiGAN \cite{kong2020hifi} is the typical GAN-based method, which utilizes the multi-receptive field fusion (MRF) module for better performance. Avocodo \cite{bak2023avocodo} jointly optimize a sub-band discriminator and a collaborative multi-band discriminator to alleviate unintended artifacts. HiFTNet \cite{li2023hiftnet} and SiFi-GAN \cite{yoneyama2023source} use the pitch-related information to improve the performance. EVA-GAN \cite{liao2024eva} adopts the ConvNeXt-based architecture to augment the context window and directly predicts the full-band signal. In particular, BigVGAN \cite{leebigvgan} achieves the state-of-the-art synthesis quality of speech with the periodic activations and anti-aliased multi-periodicity composition (AMP) module in the generator. Although GAN-based direct generation methods achieve the high fidelity, the inductive bias in time-frequency mel-spectrogram is not well utilized, which degrades the synthesized waveform quality to a certain extent.

\noindent{\bf iSTFT-based generation.} Another explored neural vocoder is the iSTFT-based network architecture. These systems usually reconstruct the waveform by parameterizing the model to predict full-band Fourier spectral coefficients, i.e., phase and magnitude components. The iSTFTNET \cite{kaneko2022istftnet} and iSTFTNET2 \cite{kaneko2023istftnet2} are a series of researches that make some modifications of HiFiGAN. Some upsample blocks with transposed convolutions are replaced with the inverse STFT in order to return Fourier spectral coefficients. In addition, some iSTFT-based vocoders are explored without upsample blocks. HiNet \cite{ai2020neural} utilizes an amplitude spectrum predictor (ASP) to predict amplitude and an NSF-based \cite{wang2019neural} phase spectrum predictor (PSP) for the phase prediction. APNet \cite{ai2023apnet} and APNet2 \cite{du2023apnet2} design the ASP and PSP modules as parallel structures. Furthermore, Vocos \cite{siuzdak2023vocos} treats the magnitude and phase predictions as a whole block, which employs ConvNeXt \cite{liu2022convnet} blocks to predict them simultaneously. All the aforementioned methods use the inverse STFT to reconstruct the full-band waveform. Unfortunately, iSTFT-based vocoders still predict Fourier spectral coefficients in a black box without explicitly supervision.

\noindent{\bf Phase loss.} The phase information is also an important part of the speech signal. Therefore, some methods learn the phase information to expect the performance improving of vocoders. PhaseAug \cite{lee2023phaseaug} arbitrarily rotates the phase of each frequency bin to learn the one-to-many relationship of speech generation. FA-GAN \cite{shen2024fa} uses the multi-resolution real and imaginary losses to learn the phase information. Additionally, APNet \cite{ai2023apnet} and APNet2 \cite{du2023apnet2} also design the explicit phase losses for phase modeling. However, these methods with supervised phase losses ignore the weight problem as described in Figure~\ref{ss}, resulting in the suboptimal‌ performance \cite{ai2023apnet,du2023apnet2}.

\section{Method}
\label{format}

\sloppy
In this section, our proposed SCNet architecture will be introduced. To begin with, we introduce the overview architecture of the proposed model. Next, we provide detailed introductions of the subband condition network, i.e., CondNet, and the magnitude-aware anti-wrapping phase loss. Finally, we will introduce the training objectives.

\subsection{Overview}
\label{view}

\sloppy
As illustrated in Figure~\ref{frmw}, the proposed model is composed of the backbone network and the conditional network. The backbone network is the standard iSTFTNET except that the Leaky ReLU activation is replaced with the Snake activation function \cite{ziyin2020neural}. The backbone branch predicts the magnitude and phase and utilizes the inverse short-time Fourier transform (iSTFT) to reconstruct the final full-band waveform. In contrast, the conditional network is designed to predict the subband waveform that is further transformed into the frequency domain with the STFT operator. Then, with the help of two coupling blocks, these prior frequency information is integrated into the corresponding layers of the backbone network. Two networks are jointly optimized during the training process.

\begin{figure}[t]
\centering
\includegraphics[width=0.21\textwidth]{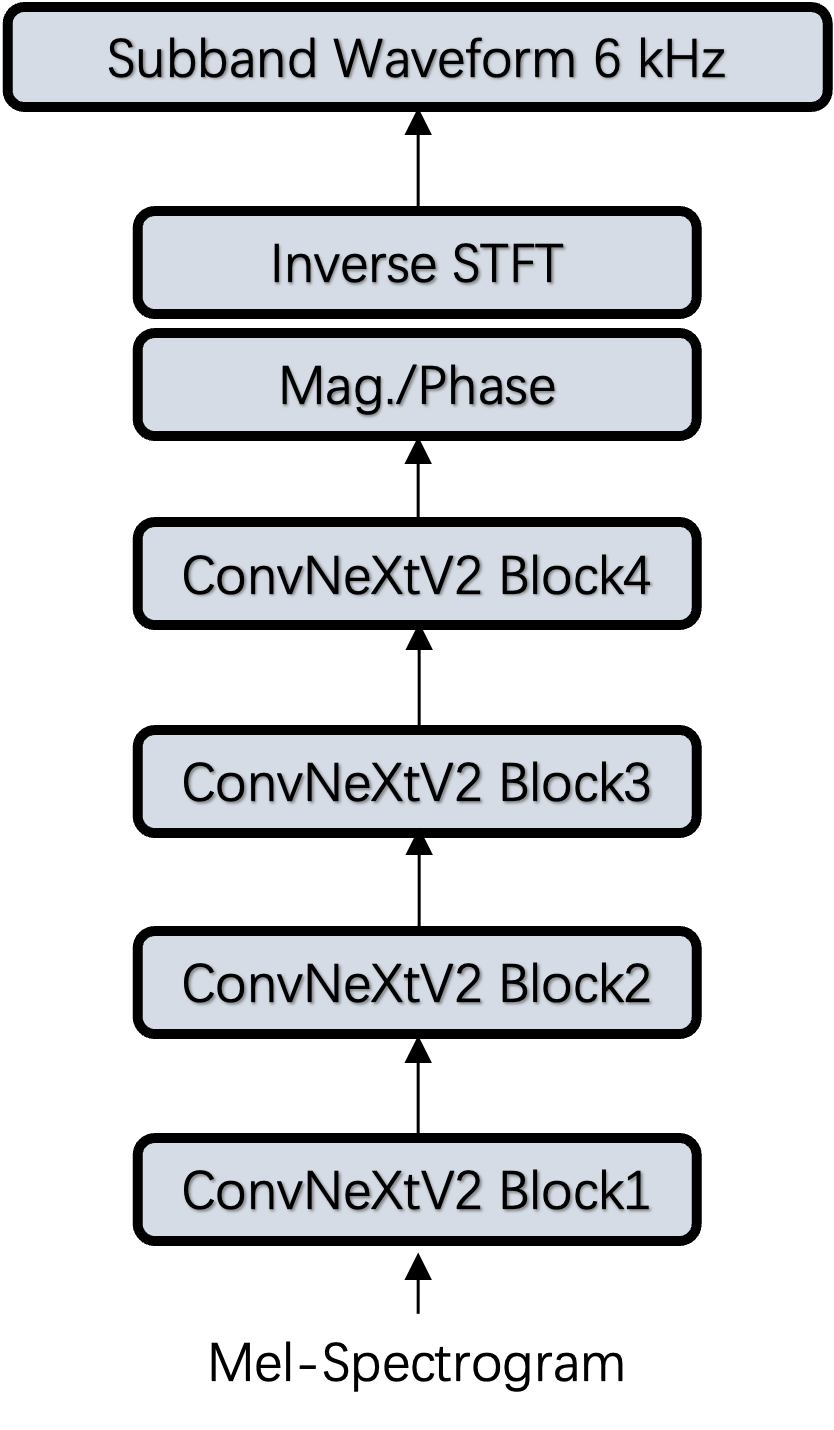}
\caption{Illustration of the subband waveform generation.}
\label{a2}
\end{figure}

\subsection{CondNet architecture}
\label{cond}

\sloppy
As the black box GAN-based generator network is lack of the guidance of the initial condition knowledge, we propose a subband conditional network, i.e., CondNet, as shown in Figure~\ref{frmw} to compensate the information loss. First, we take the mel-spectrogram representation as input and utilize the ConvNeXtV2 \cite{woo2023convnext} blocks to simultaneously predict the magnitude and phase components in one low-frequency subband. We employ the exponential function to represent the magnitude and apply the cosine and sine to denote the real and imaginary parts, respectively. Then, these complex-valued Fourier spectral coefficients are used to convert into the time domain with the inverse STFT operator. Specifically, as shown in Figure~\ref{a2}, we first use four ConvNeXtV2 blocks to predict the magnitude and phase parts of the low-frequency subband. Then, we transform the predicted values as follows
\begin{equation}
\label{eq6}
\setlength{\abovedisplayskip}{6pt}
\setlength{\belowdisplayskip}{6pt}
\hat{M} = exp\left ( \hat{m}  \right ), x = cos\left ( \hat{p}  \right ), y = sin\left ( \hat{p}  \right ),
\end{equation}
where $ \hat{m} $ and $ \hat{p} $ is the predicted magnitude and phase, respectively. Therefore, the complex-valued coefficient can be represented as $ \hat{M} \cdot \left ( x + jy \right ) $. Finally, we use the inverse STFT with the hop size 64 to generate the subband waveform with the sampling rate of 6 kHz experimentally. 

\sloppy
In order to align the output of the backbone network, i.e., magnitude and phase, we also adopt the STFT operator to obtain subband frequency representations. Next, we couple these aligned representations into corresponding layers of the backbone network. Specifically, we first employ the STFT with the hop size 4 and 1. Then, two output frequency representations are inputed into the coupling block1 and block2, respectively. Finally, we directly add these frequency domain representations to each upsample layer output respectively as the prior condition knowledge to guide the final full-band signal learning. Additionally, as shown in the right side of Figure~\ref{frmw}, the coupling block only contains the convolution layer and the Snake activation function. Notably, for the first convolution layer, we adopt the stride 2 and 1 for the first and second coupling blocks, respectively.

\subsection{Magnitude-aware phase loss}
\label{loss}
\sloppy
We first define the phase $ \theta $ calculation formula as follows
\begin{equation}
\label{eq1}
\setlength{\abovedisplayskip}{6pt}
\setlength{\belowdisplayskip}{6pt}
\theta = \arctan\left (  \frac{I}{R} \right ) - \frac{\pi }{2}\cdot Sgn\left ( I \right ) \cdot \left [ Sgn\left ( R \right ) - 1 \right ], 
\end{equation}
where $ R $ and $ I $ refer to the real part and imaginary part, respectively. When $ x \ge 0 $, $ Sgn\left ( x \right ) = 1 $;  otherwise, $ Sgn\left ( x \right ) = -1 $. Therefore, the formula restricts the phase value to the principal value interval $ \left ( -\pi, \pi   \right ] $.

\begin{figure}[t]
\includegraphics[width=\columnwidth]{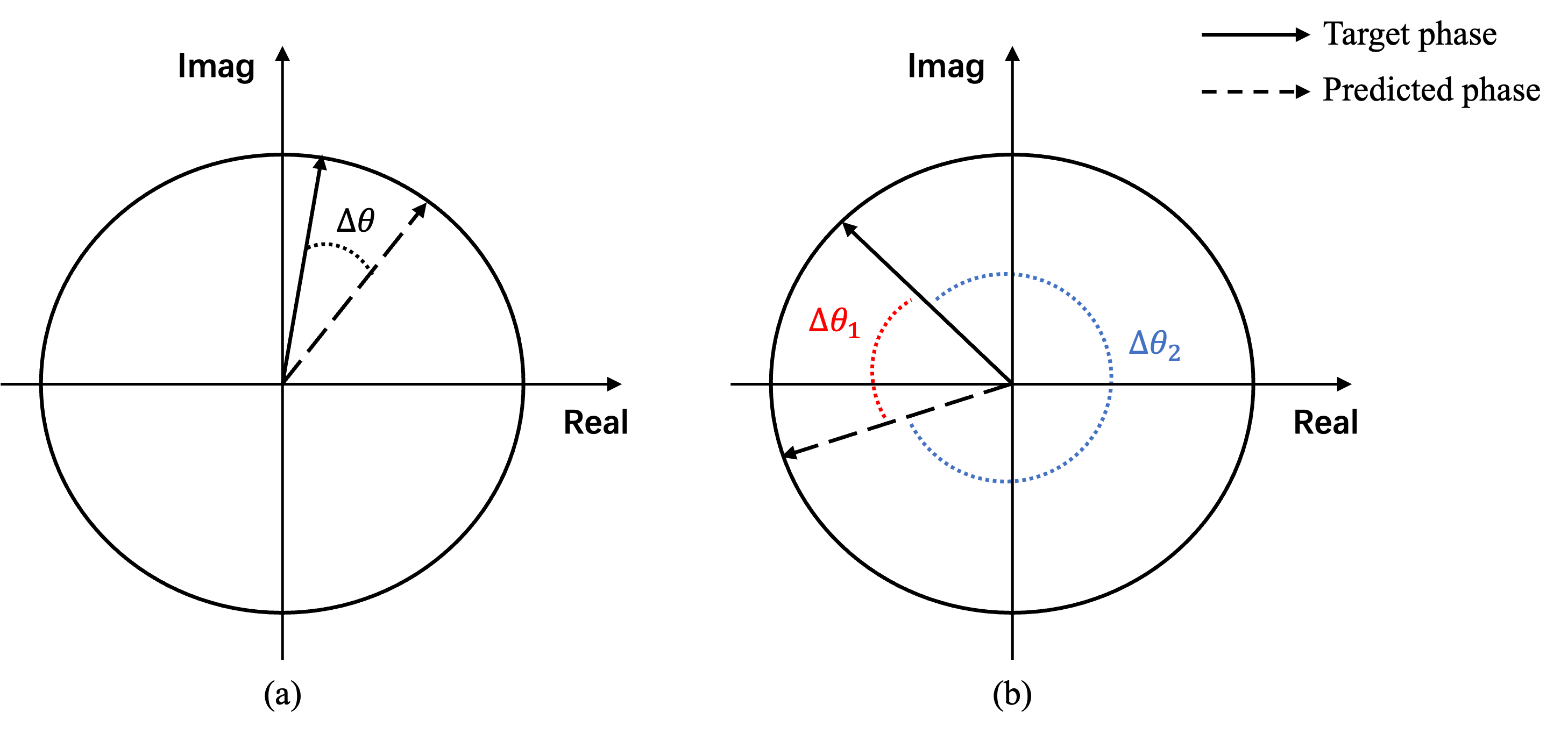}
\caption{An illustration used to explain the phase error calculation issue caused by phase wrapping.}
\label{wrp}
\end{figure}

\sloppy
Notably, the phase always has the wrapping property, resulting in the incorrect error evaluation between the raw phase $ \theta $ and the predicted phase $ \hat{\theta} $ when directly using the L1 loss or mean square error loss \cite{ai2023apnet}. As shown in Figure~\ref{wrp}, in the subfigure (a), the absolute phase error $ \Delta \theta = \left | \hat{\theta } - \theta  \right | $ is the true error. However, in the subfigure (b), the absolute error $ \Delta \theta _{2} $ is not the true error. Due to the phase wrapping issue, phase values can pass through the interval boundary between $ -\pi $ and $ \pi $. This means that the wrapping phase error $ \Delta \theta_{1} = 2 \pi - \left | \hat{\theta } - \theta  \right | $ is the true error. Therefore, we can define the calculation formula of the true phase error as follows
\begin{equation}
\label{eq2}
\setlength{\abovedisplayskip}{6pt}
\setlength{\belowdisplayskip}{6pt}
\Delta \theta = \min \left \{ \left | \hat{\theta } - \theta \right |, 2\pi -\left | \hat{\theta } - \theta \right | \right \}.
\end{equation}

\sloppy
Obviously, the true phase error $ \Delta \theta $ is restricted within the interval $ \left ( -\pi, \pi   \right ] $. According to \cite{ai2023apnet}, an ideal anti-wrapping phase loss is required to satisfy the parity, periodicity and monotonicity properties. Inspired by this rule, we design a phase loss that is defined as follows
\begin{equation}
\label{eq3}
\setlength{\abovedisplayskip}{6pt}
\setlength{\belowdisplayskip}{6pt}
\mathcal{L}_{pha} = \sin ^{2} \left ( \frac{\Delta \theta }{2}  \right ),
\end{equation}
where $ \Delta \theta $ is the definition in Equation \ref{eq2}.  Obviously, this loss is suitable for the definition of the anti-wrapping phase loss according to the above three properties. However, as shown in Figure~\ref{ss}, the loss equally considers phase errors of all time-frequency bins, which degrades the overall performance. Therefore, we design a novel magnitude-aware anti-wrapping phase loss that takes the target magnitude into account
\begin{equation}
\label{eq4}
\setlength{\abovedisplayskip}{6pt}
\setlength{\belowdisplayskip}{6pt}
\mathcal{L}_{pha} = M\cdot \sin ^{2} \left ( \frac{\Delta \theta }{2}  \right ),
\end{equation}
where $ M $ is the magnitude of the target signal. Notably in this way, the small phase error with the large magnitude can be considered as more attention. 

\subsection{Training objectives}
\label{obj}

\sloppy
In our paper, we also use the MPD and MRD as our discriminators that are consistent with BigVGAN. The training objectives are composed of adversarial loss, feature matching loss and reconstruction loss. Specifically, the least-squares (LSGAN) adversarial loss \cite{mao2017least} and feature matching $ L1 $ loss \cite{Larsen2016} are the same as BigVGAN and are only applied to the backbone network. The reconstruction loss comprises the losses on mel-spectrogram and phase, which is used for both backbone and conditional networks as follows
\begin{equation}
\label{eq5}
\setlength{\abovedisplayskip}{6pt}
\setlength{\belowdisplayskip}{6pt}
\mathcal{L}_{R} = \lambda _{m} \left ( \mathcal{L}_{mel}^{full}  + \mathcal{L}_{mel}^{sub} \right ) +\lambda _{pha} \left ( \mathcal{L}_{pha}^{full}  + \mathcal{L}_{pha}^{sub} \right )
\end{equation}
where $ \mathcal{L}_{*}^{full} $ and $ \mathcal{L}_{*}^{sub} $ refer to losses of backbone and conditional networks, respectively. $ \mathcal{L}_{mel}^{*} $ is the $ L1 $ regression loss between the predicted and real mel-spectrograms identically as BigVGAN and $ \mathcal{L}_{pha}^{*} $ is the calculation formula in Equation \ref{eq4}, which evaluates the phase errors. Additionally, the scalar weights $ \lambda _{m} = 45 $ and $ \lambda _{pha} = 45 $.

\section{Experiments}
\label{exp}

\subsection{Datasets}
\label{data}

\sloppy
We utilize the \verb+train-clean-100+ dataset from LibriTTS \cite{zen2019libritts} with the sampling rate of 24 kHz for training. For each speech sample, the number of Mel bins is set as 80 and the frequency range is set as [0, 12] kHz. All samples extract log mel-spectrograms and apply normalization. We construct two datasets for testing. Specifically, 500 utterances are randomly selected from the \verb+train-clean-100+ dataset and the remaining utterances for training. This test dataset has the same distribution with the training dataset, called in-domain (ID) dataset. To evaluate the generalization ability of the trained model for unseen speakers, we create another 500-utterance test dataset. Samples are randomly chosen from the VCTK dataset \cite{yamagishi2019cstr}, called out-of-domain (OD) dataset, which has the out-of-domain distribution compared to the training dataset.

\begin{table}[h]
\renewcommand{\arraystretch}{1.2}
\centering
\caption{Hyperparameters of the proposed SCNet.}
\label{ta1}
\resizebox{0.48\textwidth}{!}{
\setlength{\tabcolsep}{1.3mm}{
\begin{tabular}{*{3}{c}}
  \toprule
  {{\bf SCNet}} & {{\bf Module}} & {{\bf Hyperparameter}} \\
  \midrule
  \multirow{8}*{{Backbone Network}}  & {{Upsampling Ratio}} & [8, 8] \\  
   & {{Upsampling Kernel Size}} & [16, 16] \\
   & {{Upsampling Initial Channel}} & 512 \\
   & {{Upsampling Final Channel}} & 128 \\
   & {{ResBlock Kernel Size}} & [3, 7, 11] \\
   & {{ResBlock Dilation Size}} & 3$\times$[1,3,5] \\
   & {{Final Conv Output Channel}} & 18 \\
   & {{ISTFT N\_FFT and Hop Size}} & (16, 4) \\
  \midrule
   \multirow{4}*{{Condition Network}} & {{Input Channel}} & 256 \\  
   & {{Intermediate Channel}} & 768 \\
   & {{ConvNeXtV2 Layer Num}} & 4 \\
   & {{ISTFT N\_FFT and Hop Size}} & (256, 64) \\
  \midrule
   \multirow{6}*{{Coupling Network}} & {{Block1 Kernel and Stride}} & (4, 2) \\  
   & {{Block2 Kernel and Stride}} & (1, 1) \\
   & {{Conv Kernel Size}} & [7, 11] \\
   & {{Conv Dilation Size}} & [1,3] \\
   & {{Block1 Output Channel}} & 256 \\
   & {{Block2 Output Channel}} & 128 \\
   \bottomrule
\end{tabular}}}
\end{table}

\begin{table*}[t]
\renewcommand{\arraystretch}{1.2}
\centering
\caption{The experiment results of different vocoders on speech datasets in terms of in-domain and out-of-domain samples. The best results are listed in bold.}
\label{tab1}
\resizebox{1.0\textwidth}{!}{
\setlength{\tabcolsep}{3mm}{
\begin{tabular}{*{13}{c}}
  \toprule
  \multirow{2}*{{\bf Method}}  & \multicolumn{2}{c}{{\bf PESQ $\uparrow$}} & \multicolumn{2}{c}{{\bf M-STFT $\downarrow$}} & \multicolumn{2}{c}{{\bf Periodicity $\downarrow$}} & \multicolumn{2}{c}{{\bf V/UV F1 $\uparrow$}} & \multicolumn{2}{c}{{\bf Pitch $\downarrow$}} & \multicolumn{2}{c}{{\bf MOS $\uparrow$}}\\  
  \cmidrule(lr){2-3}\cmidrule(lr){4-5}\cmidrule(lr){6-7}\cmidrule(lr){8-9}\cmidrule(lr){10-11}\cmidrule(lr){12-13}
  & {{\bf ID}} & {{\bf OD}} & {{\bf ID}} & {{\bf OD}} & {{\bf ID}} & {{\bf OD}} & {{\bf ID}} & {{\bf OD}} & {{\bf ID}} & {{\bf OD}} & {{\bf ID}} & {{\bf OD}} \\
  \midrule
   {Ground Truth }& 4.50 & 4.50 & 0.00 & 0.00  & - & - & - & - & - & - & 4.59$\pm$0.12 & 4.51$\pm$0.11 \\
   \midrule
  {HiFiGAN}& 3.23 & 3.07 & 0.903 & 0.993 & 0.112 & 0.107 & 0.957 & 0.931 & 37.01 & 43.15 & 3.99$\pm$0.12 & 3.93$\pm$0.14  \\
  {iSTFTNet}& 3.10 & 3.01 & 0.934 & 1.031 & 0.114 & 0.101 & 0.957 & 0.939 & 39.69 & 44.69 & 3.96$\pm$0.09 & 3.97$\pm$0.11 \\
  {HiFTNet}& 3.53 & 3.47 & 0.830 & 0.906 & 0.095 & 0.089 & 0.964 & 0.947 & 27.26 & 30.31 & 4.06$\pm$0.11 & 4.01$\pm$0.15 \\
  {Vocos}& 3.48 & 3.40 & 0.837 & 0.890 & 0.092 & 0.083 & 0.966 & 0.952 & 27.70 & 30.49 & 4.02$\pm$0.09 & 3.99$\pm$0.12\\
  {BigVGAN}& 3.74 & 3.66 & 0.796 & 0.881 & 0.095 & 0.090 & 0.964 & 0.942 & 28.53 & 33.96 & 4.11$\pm$0.10 & 4.07$\pm$0.11 \\
  {SCNet}& {\bf 4.02} & {\bf 3.78} & {\bf 0.740} & {\bf 0.839} & {\bf 0.070} & {\bf 0.073} & {\bf 0.976} & {\bf 0.959} & {\bf 20.11} & {\bf 24.75} & {\bf 4.21$\pm$0.09} & {\bf 4.14$\pm$0.10} \\
  \bottomrule
\end{tabular}}}
\end{table*}

\begin{table*}[t]
\renewcommand{\arraystretch}{1.2}
\centering
\caption{The objective experiment results on LibriTTS dev subsets. For the pitch metric,  we use the official BigVGAN checkpoints with 5M training steps. Other objective metrics of models* are reported by BigVGAN \cite{leebigvgan} with 1M training steps.} 
\label{tab11}
\resizebox{1.0\textwidth}{!}{
\setlength{\tabcolsep}{6mm}{
\begin{tabular}{*{7}{c}}
  \toprule
  {{\bf Method}} & {{\bf Params (M)}} & {{\bf PESQ $\uparrow$}} & {{\bf M-STFT $\downarrow$}} & {{\bf Periodicity $\downarrow$}} & {{\bf V/UV F1 $\uparrow$}} &{{\bf Pitch $\downarrow$}} \\  
  \midrule
  {BigVGAN-base*}& 14.01 & 3.519 & 0.8788 & 0.1287 & 0.9459 & {\bf 24.432} \\
  {BigVGAN*}& 112.4 & {\bf 4.027} & {\bf 0.7997} & 0.1018 & 0.9598 & 25.651 \\
  {SCNet (1M steps)}& 15.86 & 3.881 & 0.8278 & 0.1007 & 0.9580 & 26.031  \\ 
  {SCNet (2M steps)}& 15.86 & 4.007 & 0.8070 & {\bf 0.0950} & {\bf 0.9604} & 25.533  \\ 
  \bottomrule
\end{tabular}}}
\end{table*}

\begin{table}[t]
\renewcommand{\arraystretch}{1.2}
\centering
\caption{The MOS results of different vocoders. Acoustic features are generated from CosyVoice.}
\label{tt1}
\resizebox{0.4\textwidth}{!}{
\setlength{\tabcolsep}{11mm}{
\begin{tabular}{*{2}{c}}
  \toprule
   {{\bf Method}} & {{\bf MOS}}  \\  
  \midrule
 {HiFiGAN}& 3.70$\pm$0.12  \\ 
 {iSTFTNet}& 3.76$\pm$0.10 \\
 {HiFTNet}& 3.88$\pm$0.09 \\
 {Vocos}& 3.91$\pm$0.12 \\
 {BigVGAN}& 4.01$\pm$0.08 \\
 {SCNet}& {\bf 4.09$\pm$0.10} \\
 \bottomrule
\end{tabular}}}
\end{table}

\subsection{Training setups}
\label{tb}
 
\sloppy
\noindent{\bf Model architecture.} In SCNet, we predict the subband waveform with the sampling rate of 6 kHz. For the backbone branch, we use two upsample blocks, each contains the transposed convolution layer with the kernel size (16, 16). These upsample blocks achieve 64x upsampling, where up-factor is (8, 8). The multi-receptive field fusion (MRF) modules have the same configurations with BigVGAN. The output channels of two upsample blocks are 256 and 128, respectively. In addition, the output channels of the first and last convolution layers are 512 and 18, respectively. For CondNet, we adopt 4 ConvNeXtV2 blocks to predict the subband Fourier spectral coefficients and the input channel of the first block is converted to 256. The intermediate dimension is 768 and the output channels of the final convolution layers are both 129 for phase and magnitude predicting. Then, we use the inverse STFT with the hop size 64 to generate the low-frequency subband waveform. Furthermore, we employ two STFT modules with the hop size 4 and 1, respectively. In the coupling block, there are one normal convolution layer and two dilated convolution layers with the kernel sizes (7, 11) and dilations (1, 3). The output channels of two coupling blocks are 256 and 128, respectively. For the backbone network, we use the inverse STFT with the frame length 16 and frame shift 4 to generate the final waveform. We present the hyperparameter details of SCNet in Table~\ref{ta1}. Additionally, the source code of SCNet is available at the website\footnote{The source code is available at \url{https://github.com/vspeech/SCNet}}.

\sloppy
\noindent{\bf Training.} For the magnitude-aware phase loss, FFT size, window size and hop size are set as (1024, 1024, 256) for the backbone and (256, 256, 64) for CondNet, respectively. Notably, 24 kHz is the original sampling rate and 6 kHz is applied for the predicted subband waveform in CondNet. Therefore for the subband mel-spectrogram loss, the number of Mel bins is set as 20. Furthermore, we randomly intercept a segment size of 24576 for each speech sample and apply the batch size as 16 during training process. The weight normalization is employed for all modules. The initial learning rate of generator and discriminator is set as 2e-4 with an exponential decay rate of 0.999 and the model is optimized using AdamW optimizer \cite{kingma2014adam} with betas (0.8, 0.99). Finally, we train all models up to 1M steps on an NVIDIA A100 GPU. For all ablation experiments, we only train related models for 0.5M steps. 

\subsection{Baselines and evaluations}
\label{eva}

\sloppy
\noindent{\bf Baselines.} Two direct waveform generation methods (HiFiGAN\footnote{\url{https://github.com/jik876/hifi-gan}} \cite{kong2020hifi}, BigVGAN\footnote{\url{https://github.com/NVIDIA/BigVGAN}} \cite{leebigvgan}) and three iSTFT-based generation methods (iSTFTNet \cite{kaneko2022istftnet}, HiFTNet\footnote{\url{https://github.com/yl4579/HiFTNet}} \cite{li2023hiftnet} and Vocos\footnote{\url{https://github.com/gemelo-ai/vocos}} \cite{siuzdak2023vocos}) are used as baseline models. We retrain all baselines using public official codes other than iSTFTNet. In addition, we also utilize the unofficial implementation for iSTFTNet\footnote{\url{https://github.com/rishikksh20/iSTFTNet-pytorch}} training. Notably, we train all compared vocoders on the same configurations as mentioned in SCNet.

\begin{table*}[t]
\renewcommand{\arraystretch}{1.2}
\centering
\caption{The detailed training and synthesis speeds of different vocoders. Model footprint and training days of 1M steps are listed. The synthesis speed is measured on an NVIDIA V100 GPU.}
\label{tab4}
\resizebox{1.0\textwidth}{!}{
\setlength{\tabcolsep}{4.1mm}{
\begin{tabular}{*{8}{c}}
  \toprule
  {{\bf Method}}  & {{\bf Type}} & {{\bf Params (M)}} & {{\bf Segment}} & {{\bf Batch size}} & {{\bf Training days}} & {{\bf Training memory}} & {{\bf Syn. speed}} \\  
  \midrule
 {HiFiGAN}& direct & 14.01 & 24576 & 16 & 5.4  & 19GB & 112.81 \\ 
 {BigVGAN}& direct & 112.4 & 24576 & 16 & 11.1 & 72GB & 41.48 \\ 
  {iSTFTNet}& iSTFT & 13.30 & 24576 & 16 & 4.2 & 16GB & 184.65 \\ 
  {HiFTNet}& iSTFT & 21.42 & 24576 & 16 & 4.8 & 24GB & 89.50  \\ 
  {Vocos}& iSTFT & 13.50 & 24576 & 16 & 2.6 & 13GB & 609.01 \\ 
  {SCNet}& iSTFT & 15.86 & 24576 & 16 & 3.6 & 15GB & 145.67 \\ 
 \bottomrule
\end{tabular}}}
\end{table*}

\begin{table*}[t]
\renewcommand{\arraystretch}{1.2}
\centering
\caption{The results of conditional network ablation experiments on speech dataset in terms of in-domain and out-of-domain samples with 0.5M training steps. The best results are listed in bold.}
\label{tab2}
\resizebox{1.0\textwidth}{!}{
\setlength{\tabcolsep}{4.5mm}{
\begin{tabular}{*{11}{c}}
  \toprule
  \multirow{2}*{{\bf Method}} & \multicolumn{2}{c}{{\bf PESQ $\uparrow$}} & \multicolumn{2}{c}{{\bf M-STFT $\downarrow$}} & \multicolumn{2}{c}{{\bf Periodicity $\downarrow$}} & \multicolumn{2}{c}{{\bf V/UV F1 $\uparrow$}} & \multicolumn{2}{c}{{\bf Pitch $\downarrow$}}\\  
  \cmidrule(lr){2-3}\cmidrule(lr){4-5}\cmidrule(lr){6-7}\cmidrule(lr){8-9}\cmidrule(lr){10-11}
  & {{\bf ID}} & {{\bf OD}} & {{\bf ID}} & {{\bf OD}} & {{\bf ID}} & {{\bf OD}} & {{\bf ID}} & {{\bf OD}} & {{\bf ID}} & {{\bf OD}} \\
  \midrule
  {SCNet}& {\bf 3.90} & {\bf 3.67} & {\bf 0.768} & {\bf 0.867} & {\bf 0.078} & {\bf 0.079} & {\bf 0.972} & {\bf 0.954} & {\bf 21.86} & {\bf 27.28}  \\
  {w/ full-band}& 3.82 & 3.54 & 0.789 & 0.896 & 0.084 & 0.083 & 0.970 & 0.953 & 23.73 & 30.11  \\
  {w/ time cond}& 3.88 & 3.56 & 0.772 & 0.882 & 0.079 & 0.081 & 0.972 & 0.954 & 21.89 & 27.96 \\
  {w/o CondNet}& 3.31 & 3.06 & 0.874 & 0.989 & 0.112 & 0.100 & 0.958 & 0.940 & 34.34 & 40.06  \\
  \bottomrule
\end{tabular}}}
\end{table*}

\sloppy
\noindent{\bf Evaluations.} In our experiments, we utilize both objective and subjective evaluations for our proposed model and baselines. For the objective evaluations, we adopt 5 different metrics, i.e., the Perceptual Evaluation of Speech Quality (PESQ) \cite{rix2001perceptual} with 16 kHz wide-band version\footnote{\url{https://github.com/ludlows/python-pesq.}}, Multi-resolution STFT (M-STFT) \cite{yamamoto2020parallel} that measures the difference of spectral distance with multiple resolutions\footnote{We use the open-source tool from Auraloss \cite{ste2020aura}.} and 3 pitch-related metrics. The pitch-related metrics contain Periodicity error, F1 score of voiced/unvoiced classification (V/UV F1) and pitch error using F0 Root Mean Square Error\footnote{We use an open-source code provided by CARGAN \cite{Morr2022cargan}.}. For the subjective evaluation, we rely on the crowd-sourced 5-point Mean Opinion Score (MOS) metric to estimate the speech quality and intelligibility of test datasets. Specifically, twenty raters listen to randomly chosen speech samples, and score their naturalness from 1 to 5. Score 1 indicates poor speech and score 5 denotes excellent speech. Raters are allowed to evaluate each speech sample once. To assess inference speed, an NVIDIA V100 GPU is used to evaluate the average synthesis speed for generating 500 in-domain speech samples and the xRT value that means the speed factor relative to real-time is used for speed evaluation. Value 1.0 of xRT denotes real-time speed.

\subsection{Results}
\label{res}

\subsubsection{Model performance}

\sloppy
We first evaluate the performance of our proposed SCNet model compared to the GAN-based baseline models, as illustrated in Table~\ref{tab1}. In terms of all metrics, our proposed SCNet realizes the more superior performance compared to the other baseline models that has the same level parameters (as shown in Table~\ref{tab4}). Additionally, although the previous state-of-the-art high capacity GAN-based model BigVGAN has larger parameter, i.e., 112M, our SCNet still achieves better performance compared to BigVGAN on the in-domain and out-of-domain test datasets and only uses approximately 1/8 lighter in model size compared to BigVGAN. These findings indicate that the proposed SCNet can achieve significantly superior performance compared to all GAN-based baseline models on the small-scale training dataset. Furthermore, SCNet notably exhibits consistently improved metric scores over other baselines on the out-of-domain dataset with unseen speakers, which verifies the superior generalization capability of the proposed model. 

\begin{table*}[t]
\renewcommand{\arraystretch}{1.2}
\centering
\caption{The results of phase loss ablation experiments on speech dataset in terms of in-domain and out-of-domain samples with 0.5M training steps. The best results are listed in bold.}
\label{tab3}
\resizebox{1.0\textwidth}{!}{
\setlength{\tabcolsep}{4.5mm}{
\begin{tabular}{*{11}{c}}
  \toprule
  \multirow{2}*{{\bf Method}} & \multicolumn{2}{c}{{\bf PESQ $\uparrow$}} & \multicolumn{2}{c}{{\bf M-STFT $\downarrow$}} & \multicolumn{2}{c}{{\bf Periodicity $\downarrow$}} & \multicolumn{2}{c}{{\bf V/UV F1 $\uparrow$}} & \multicolumn{2}{c}{{\bf Pitch $\downarrow$}}\\  
  \cmidrule(lr){2-3}\cmidrule(lr){4-5}\cmidrule(lr){6-7}\cmidrule(lr){8-9}\cmidrule(lr){10-11}
  & {{\bf ID}} & {{\bf OD}} & {{\bf ID}} & {{\bf OD}} & {{\bf ID}} & {{\bf OD}} & {{\bf ID}} & {{\bf OD}} & {{\bf ID}} & {{\bf OD}} \\
  \midrule
  {SCNet}& {\bf 3.90} & {\bf 3.67} & {\bf 0.768} & {\bf 0.867} & {\bf 0.078} & {\bf 0.079} & {\bf 0.972} & {\bf 0.954} & {\bf 21.86} & {\bf 27.28}  \\
  {w/o phase loss}& 3.80 & 3.49 & 0.787 & 0.886 & 0.088 & 0.085 & 0.968 & 0.951 & 24.54 & 31.10 \\
  {w/o mag weight}& 3.78 & 3.49 & 0.786 & 0.882 & 0.087 & 0.084 & 0.966 & 0.951 & 24.49 & 30.53 \\
  {w/ APNet2 loss}& 3.77 & 3.51 & 0.789 & 0.884 & 0.089 & 0.085 & 0.968 & 0.950 & 24.77 & 30.93 \\
  \bottomrule
\end{tabular}}}
\end{table*}

\begin{table*}[t]
\renewcommand{\arraystretch}{1.2}
\centering
\caption{The experiment results of different phase loss weights on speech dataset in terms of in-domain and out-of-domain samples with 0.5M training steps. The best results are listed in bold.}
\label{ta2}
\resizebox{1.0\textwidth}{!}{
\setlength{\tabcolsep}{5mm}{
\begin{tabular}{*{11}{c}}
  \toprule
  \multirow{2}*{{\bf Method}} & \multicolumn{2}{c}{{\bf PESQ $\uparrow$}} & \multicolumn{2}{c}{{\bf M-STFT $\downarrow$}} & \multicolumn{2}{c}{{\bf Periodicity $\downarrow$}} & \multicolumn{2}{c}{{\bf V/UV F1 $\uparrow$}} & \multicolumn{2}{c}{{\bf Pitch $\downarrow$}}\\  
  \cmidrule(lr){2-3}\cmidrule(lr){4-5}\cmidrule(lr){6-7}\cmidrule(lr){8-9}\cmidrule(lr){10-11}
  & {{\bf ID}} & {{\bf OD}} & {{\bf ID}} & {{\bf OD}} & {{\bf ID}} & {{\bf OD}} & {{\bf ID}} & {{\bf OD}} & {{\bf ID}} & {{\bf OD}} \\
  \midrule
  {weight 25}& 3.88 & 3.64 & {\bf 0.767} & 0.869 & 0.082 & 0.081 & 0.970 & 0.952 & 22.79 & 29.88 \\
  {weight 45}& 3.90 & {\bf 3.67} & 0.768 & {\bf 0.867} & {\bf 0.078} & {\bf 0.079} & {\bf 0.972} & 0.954 & {\bf 21.86} & 27.28 \\
  {weight 60}& {\bf 3.91} & 3.63 & 0.769 & 0.868 & 0.078 & 0.079 & 0.972 & {\bf 0.955} & 22.03 & {\bf 26.78} \\
  {weight 90}& 3.85 & 3.53 & 0.778 & 0.881 & 0.079 & 0.082 & 0.971 & 0.953 & 22.64 & 28.53 \\
  \bottomrule
\end{tabular}}}
\end{table*}

\sloppy
As the closest competitor, BigVGAN cannot reflect its ability on small-scale training data, which is also demonstrated in \cite{leebigvgan}. To further compare our model to BigVGAN on the large-scale training data, we also train our model on the \verb+train-full-960+ dataset of LibriTTS \cite{zen2019libritts} and evaluate the trained model on the \verb+dev+ subsets (\verb+dev-clean+ and \verb+dev-other+) of LibriTTS. For the sake of fairness, the number of Mel bins is set as 100 and the batch size is 32. As shown in Table~\ref{tab11}, when training 1M steps, SCNet achieves better performance compared to BigVGAN-base on most metrics (Pitch metric has a slight decline). However, there is a certain performance gap with BigVGAN. When training 2M steps, SCNet achieves the competitive performance compared to BigVGAN on all metrics. These findings indicate that our proposed method with the small-scale parameter also has the superior generalization ability on unseen speech samples.

\subsubsection{TTS experiments}
\sloppy
Generally, vocoder is an important part for text-to-speech. Therefore, we use the recent Cosyvoice \cite{du2024cosyvoice} as the TTS baseline model and randomly select 50 utterances from dev subsets of librettos \cite{zen2019libritts} dataset. For MOS evaluation, a total of twenty people participate and participants are required to evaluate each utterance once for all models. Notably, we utilize all trained models in Table~\ref{tab1} for evaluation. As shown in Table~\ref{tt1}, the quality of the generated speech using SCNet achieves the 4.09 MOS and outperforms other baseline models.

\begin{figure}[t]
\includegraphics[width=\columnwidth]{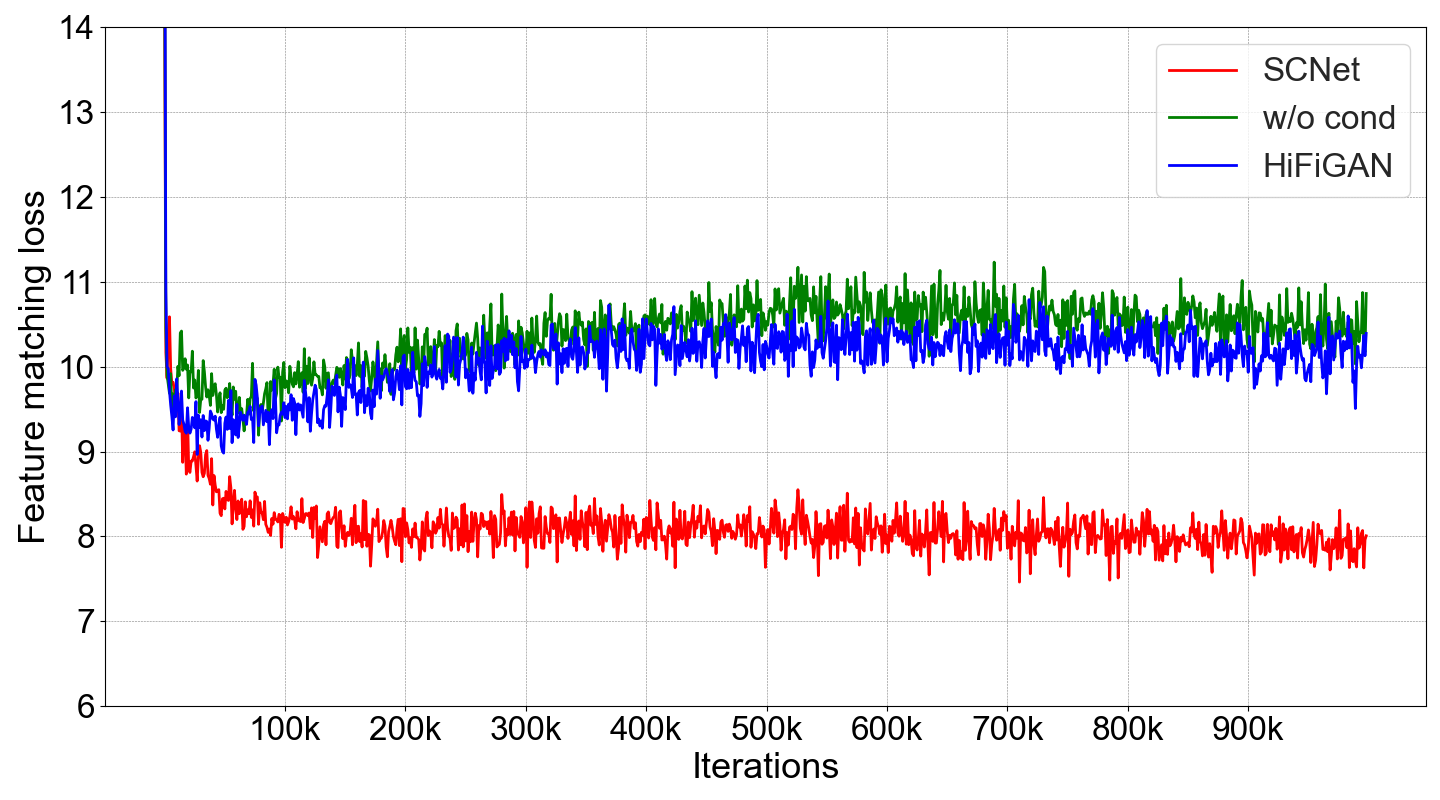}
\caption{The illustrations of feature matching loss in terms of SCNet, SCNet without CondNet and HiFiGAN. Notably, we train three models up to 1M steps with the same MPD and MRD discriminators for comparability.}
\label{a1}
\end{figure}

\subsubsection{Analysis of feature matching loss}
\sloppy
To demonstrate the effectiveness of the proposed condition network, we also train the proposed SCNet, SCNet without CondNet and HiFiGAN on \verb+train-clean-100+ dataset from LibriTTS \cite{zen2019libritts}, respectively. The same discriminators, i.e., MPD and MRD are utilized during the training process. The experiment results of different feature matching losses are shown in Figure~\ref{a1}. Notably, feature matching losses without conditional prior information (green and blue lines) gradually increase until a stable value. In contrast, the loss of SCNet with the conditional prior information (red line) gradually decreases until a stable value. The main reason is that the GAN-based vocoders are usually trained in a black-box way. The intermediate features have no guidance, which leads to the unstable feature matching loss. The condition network brings the prior knowledge to the backbone branch, which is conducive to the model training.

\subsubsection{Training and synthesis speeds}
\sloppy
For the training and synthesis speeds, we compare our proposed SCNet to all baseline models. As shown in Table~\ref{tab4}, SCNet achieves fast and comparable synthesis speed than all GAN-based models. Specifically, SCNet realizes slightly faster than HiFiGAN, and approximately 4 times faster than BigVGAN. In addition, SCNet also owns the comparable synthesis speed compared to iSTFTNET. This is mainly because of the injection of the subband condition network. Furthermore, while SCNet is more than 4 times slower of the synthesis speed compared to Vocos, it achieves more superior performance than Vocos for both objective and subjective evaluations. Therefore, SCNet achieves a better balance between performance and synthesis speed. Additionally, we use the same configurations to train all models on NVIDIA A100 GPU. For the training time, baseline models other than Vocos consume more training days when training 1M steps. Notably, BigVGAN consumes more than 3 times training days compared to the SCNet. Furthermore, SCNet also utilizes the less training memory compared to other baselines other than Vocos.

\subsubsection{Ablation study}

\sloppy
\noindent{\bf CondNet architecture.} To demonstrate the effectiveness of the proposed CondNet, we also conduct some ablation experiments of the proposed condition network. Experiment results of the condition network ablation are shown in Table~\ref{tab2}. For the CondNet, we predict the full-band waveform rather than subband waveform. This means that the larger frame shift, i.e., 256, are required for full-band waveform prediction. This substitution leads to the overall performance decline compared to the original version, especially for PESQ and Pitch metrics on out-of-domain speech dataset. This finding indicates that the predicted subband waveform contributes to better speech quality and the fundamental frequency learning. Additionally, omitting two STFT operators between the predicted subband waveform and coupling blocks (w/ time cond) also results in the slightly degraded quality. The main reason is that the prior information of the time domain misaligns the final output of the frequency domain, i.e., magnitude and phase. Notably, completely omitting the conditional network leads to a dramatical degradation compared to the original version in terms of all metrics. The prior information from the condition network provides better initial guidance and contributes to the fine-grained waveform generation. This finding also demonstrates the importance of the proposed condition network. 

\begin{figure*}[t]
\begin{minipage}[b]{1.0\linewidth}
  \centering
  \centerline{\includegraphics[width=13cm]{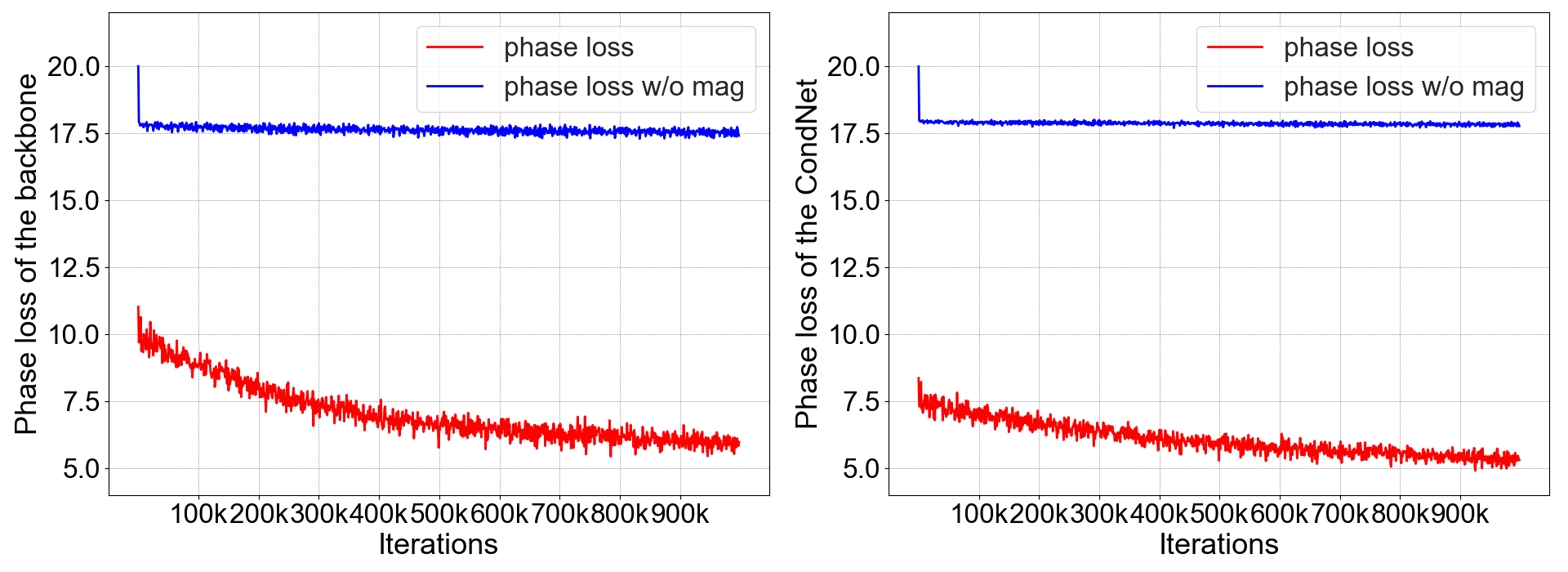}}
 \end{minipage}
\caption{The illustrations of the proposed phase loss with or without the magnitude weights. We train up to 1M steps and present the phase losses of the backbone and condition networks.}
\label{ploss}
\end{figure*}

\sloppy
\noindent{\bf Magnitude-aware phase loss.} We also conduct ablation studies in terms of the phase losses ablated. The objective experiment results are shown in Table~\ref{tab3}. Specifically, we totally drop the proposed magnitude-aware anti-wrapping phase loss during the training process. The results are shown in the second row in Table~\ref{tab3}. This manipulation results in performance declines to a certain extent, especially for the out-of-domain speech dataset, which verifies that the magnitude-aware anti-wrapping phase loss can be effectively used as a complementary view of the magnitude and further facilitate the fine-grained reconstruction of waveform. Furthermore, we directly omit the target magnitude weight in the phase loss, which results in the similar performance compared to the item without the phase loss. This finding indicates that using the phase loss in Equation~\ref{eq3} cannot contribute to the performance improvement. Therefore, the designed phase loss can benefit from the magnitude-aware weight. Additionally, we replace our proposed phase loss with the phase loss in APNet2 \cite{du2023apnet2}, which also results in performance declines. Moreover, as shown in Figure~\ref{ploss}, we draw curves of phase losses with and without the magnitude-aware weight in terms of the backbone and condition networks, respectively. Notably, omitting the magnitude-aware weight results in the non-convergence issue of the phase loss. In contrast, the proposed magnitude-aware phase loss can quickly decline during the training process, which indicates that the magnitude-aware weight can contribute to the stable model optimization. 

\sloppy
As shown in Table~\ref{ta2}, we also conduct experiments of different phase loss weights and train these models up to 0.5M steps. Note that different phase weights can achieve the approximate results, which indicates the robustness to this parameter. Notably, removing the phase loss will result in the significant performance degradation (as shown in Tabel~\ref{tab3}), which also demonstrates the effectiveness of the proposed magnitude-aware phase loss. In our experiments, we set the weight as 45. 

\begin{table}[t]
\renewcommand{\arraystretch}{1.2}
\centering
\caption{The objective experiment results on the out-of-domain speech dataset with different pre-trained subband condition networks. 0.3M and 1M refer to the training steps of the subband condition network, respectively.}
\label{tabs}
\resizebox{0.48\textwidth}{!}{
\setlength{\tabcolsep}{10mm}{
\begin{tabular}{*{3}{c}}
  \toprule
  \multirow{2}*{{\bf Metrics}} & \multicolumn{2}{c}{{\bf Subband Steps}} \\  
  \cmidrule(lr){2-3}
  & {{\bf 0.3M}} & {{\bf 1M}} \\
  \midrule
  {PESQ}& 3.34 & {\bf 3.69} \\
  {M-STFT}& 0.931 & {\bf 0.871} \\
  {Periodicity}&  0.093 & {\bf 0.081} \\
  {V/UV F1}& 0.945 & {\bf 0.951} \\
  {Pitch}& 37.02 & {\bf 29.21} \\
  \bottomrule
\end{tabular}}}
\end{table}

\sloppy
\noindent{\bf Impacts about the quality of subband waveform.} To further demonstrate the contribute of the proposed subband condition network, we conduct the related experiments and analyze the impacts of the generated subband waveform on the final output. Specifically, we first select the subband condition network with different training steps, i.e., 0.3M and 1M. Then, we initialize the CondNet utilizing the pre-trained parts and freeze them to retrain the remaining parts of SCNet up to 0.5M training steps, respectively. As shown in Table~\ref{tabs}, SCNet with the fixed 1M-step subband condition network significantly outperforms that with the fixed 0.3M-step subband condition network. Notably, the better quality of the subband waveform provides better prior guidance to the backbone network, which makes the generator not entirely a black box and contributes to better performance of the final outputs.

\section{Limitations}
\label{limi}
\sloppy
While the proposed SCNet learns the prior condition information and trains with the magnitude-aware phase loss, contributing to better performance, it still remains several limitations. The SCNet has slower synthesis speed compared to Vocos. Perhaps more flexible framework combining the condition network with the backbone branch can achieve faster synthesis speed. Additionally, while SCNet has the superior performance for speech generation, a universal vocoder for speech, audio and music generation has not been verified. We will explore these limitations in the future research.

\section{Conclusion}
\label{con}
\sloppy
In this paper, we propose a dual-branch neural vocoder with the magnitude-aware anti-wrapping phase loss. Specifically in our model, the subband prediction network and iSTFT-based backbone branch are integrated into a framework. We propose the CondNet to predict the subband waveform and further use the STFT operator to generate the aligned representations of the frequency domain, which provides the prior information and simplifies the difficulty of model training. Moreover, to more correctly learn the phase structures in speech, we design a novel magnitude-aware anti-wrapping phase loss, which utilizes the magnitude as the weight to train the overall network and achieves performance gains. Our experiment results also demonstrate the effectiveness of the proposed model and phase loss. In conclusion, SCNet provides a new idea for the advancement of neural vocoders. 

\section{Use of Generative AI Disclosure}
\sloppy
I hereby declare that no AI tools or language models (such as ChatGPT, DeepSeek, or similar) were used in the writing, data analysis, or any other part of this paper. All content is solely the product of my own intellectual effort and original work.

\bibliographystyle{IEEEtran}
\bibliography{mybib}

\end{document}